\documentclass[12pt]{IEEEtai} 

\usepackage{cite}
\usepackage{amsmath,amssymb,amsfonts}
\usepackage{algorithmic}
\usepackage{graphicx}
\usepackage{textcomp}
\usepackage[normalem]{ulem}

\usepackage{amsmath}
\usepackage{amsmath}   % for \mathcal and general math symbols
\usepackage{amssymb}   % for \mathbb and additional math symbols
\usepackage{amsfonts,mathrsfs}
\usepackage{bm,bbm} % \mathbbm font
\usepackage{dsfont} % \mathds font

\usepackage{amsthm} %<--- should be uncommented
\newtheorem{theorem}{Theorem}

\newtheorem{proposition}[theorem]{Proposition}

\newtheorem{lemma}[theorem]{Lemma}
\newtheorem{remark}{Remark}

\usepackage{algorithm}
\usepackage{url}
\usepackage{hyperref}

\usepackage{comment}
\usepackage{multirow}

\usepackage{booktabs}

\newcommand{\mdp}{\texttt{MDP}}
\newcommand{\dyp}{\texttt{DP}}

\newcommand{\adp}{\texttt{ADP}}

\newcommand{\td}{\texttt{TD}}
\newcommand{\lstd}{\texttt{LSTD}}

\newcommand{\rl}{\texttt{RL}}

\usepackage{circuitikz}
\usetikzlibrary{arrows,automata, quotes, calc, trees, positioning, chains, shapes.geometric, decorations.pathreplacing, decorations.pathmorphing, shapes, matrix, shapes.symbols}
\usetikzlibrary{spy,backgrounds,shadows}
\usetikzlibrary{fit,calc}

\makeatletter 
\let\myorg@bibitem\bibitem
\def\bibitem#1#2\par{%
  \@ifundefined{bibitem@#1}{%
    \myorg@bibitem{#1}#2\par
  }{%
    \begingroup
      \color{\csname bibitem@#1\endcsname}%
      \myorg@bibitem{#1}#2\par
    \endgroup
  }%
}

\onecolumn

\begin{document}
\title{\LARGE Off-Policy Temporal Difference Learning for Perturbed Markov Decision Processes: Theoretical Insights and Extensive Simulations}
\author{Ali Forootani, \IEEEmembership{Senior Member, IEEE}, Raffaele Iervolino, \IEEEmembership{Senior Member, IEEE}, Massimo Tipaldi, and Mohammad Khosravi \IEEEmembership{Member, IEEE}
\thanks{Ali Forootani is with Helmholtz Center for Environmental Research-UFZ, Permoserstrasse 15, 04318 Leipzig, Germany (\texttt{aliforootani@ieee.org/ali.forootani@ufz.de}).}
\thanks{Raffaele Iervolino is with Department of Electrical Engineering and Information Technology, University of Naples, 80125 Napoli, Italy 
 (\texttt{rafierv@unina.it}).}
\thanks{Massimo Tipaldi is with Department of Electrical and Information Engineering, Polytechnic University of Bari, 70126 Bari, Italy (\texttt{massimo.tipaldi@poliba.it}).}
\thanks{Mohammad Khosravi is with Delft Center for Systems and Control, Delft University of Technology, Mekelweg 2, 2628 CD, Delft, The Netherlands 
(\texttt{mohammad.khosravi@tudelft.nl}).}}

\maketitle

\begin{abstract}
Dynamic Programming suffers from the curse of dimensionality due to large state and action spaces, a challenge further compounded by uncertainties in the environment. To mitigate these issue, we explore an off-policy based Temporal Difference Approximate Dynamic Programming \color{black}approach that preserves contraction mapping when projecting the problem into a subspace of selected features, accounting for the probability distribution of the perturbed transition probability matrix. We further demonstrate how this Approximate Dynamic Programming \color{black}approach can be implemented as a particular variant of the Temporal Difference learning algorithm, adapted for handling perturbations. To validate our theoretical findings, we provide a numerical example using a Markov Decision Process \color{black}corresponding to a resource allocation problem.
\end{abstract}
%%%%%%%%%%%%%%%%%%%%%%%%%%%%%%%
%%%%%%%%%%%%%%%%%%%%%%%%%%%%%%%
%%%%%%%%%%%%%%%%%%%%%%%%%%%%%%%
%
\begin{IEEEkeywords}
Reinforcement Learning, Markov Decision Processes, Temporal Difference Learning, Perturbed Probability Transition Matrix.
\end{IEEEkeywords}

%%%%%%%%%%%%%%%%%%%%%%%%%%%%%%%
%%%%%%%%%%%%%%%%%%%%%%%%%%%%%%%
%%%%%%%%%%%%%%%%%%%%%%%%%%%%%%%

\section{Introduction}
\label{sec:introduction}
\IEEEPARstart{I}{n} large-scale Markov Decision Processes (\mdp s), Dynamic Programming (\dyp) \color{black}faces the curse of dimensionality, where the state and action spaces grow exponentially, making computation infeasible \cite{Forootani_IFAC}. To tackle these challenges,  Approximate Dynamic Programming (\adp) \color{black}methods are employed in the literature \cite{Forootani_cont_sys_lett, Iervolino_IJC}. \adp~techniques such as Temporal Difference (\td) use function approximations and simulations to simplify computational complexity \cite{lin2023policy, forootani2024kernel}.

%\paragraph{Literature review and Contributions\color{black}}
On-policy and off-policy \td~methods rely on three main methodologies to estimate cost functions: 
gradient-based solutions, which use stochastic gradient descent to minimize the prediction error incrementally (e.g., \cite{Sutton_emphatic_2016}, \cite{sutton2008convergent}); least-squares minimization techniques, which estimate cost functions by minimizing the least-squares error between the estimated and true returns (e.g., \cite{bertsekas2011temporal}, \cite{Forootani_IFAC}, \cite{choi2006generalized}); and probabilistic approaches, which employ Bayesian techniques to model the uncertainty in value estimates, adjusting predictions based on the confidence in the current estimates (e.g. \cite{geist2010kalman}, \cite{engel2005reinforcement}).  Off-policy \td~is a powerful technique within the broader field of \adp~that aims to estimate cost functions using data generated by a policy different from the one being improved\cite{Forootani_IFAC}. Unlike on-policy methods, which require following the same policy being evaluated, off-policy \td~learning offers greater flexibility by leveraging experience from a variety of policies or even simulated environments\cite{schulman2017proximal}. In particular, off-policy \td~approaches enable more extensive exploration of the environment, potentially resulting in better policies\cite{bertsekas2023course}. In \cite{schulmanhigh}, generalized advantage estimation was proposed, which smooths the trade-off by reducing variance in policy gradient updates, allowing more stable learning in high-dimensional control tasks. Exploration strategies, pivotal in Reinforcement Learning (\rl), were demonstrated effectively in their greedy approach within Deep \texttt{Q}-Learning to achieve human-level control in \texttt{Atari} games \cite{mnihhuman}. Further advanced exploration-exploitation trade-offs with Proximal Policy Optimization (\texttt{PPO}) was introduced in \cite{schulman2017proximal}, which uses a clipped surrogate objective to ensure stable policy updates while balancing exploration and exploitation, making it robust for continuous and discrete action spaces.\color{black}

\mdp~leverages simulations to generate experience and approximate optimal cost functions that lead to effective policies. However, inaccuracies in the model and dynamic changes in the environment can introduce challenges. If the model does not accurately represent the environment or if the environment changes over time, the effectiveness of a policy based on the model may be compromised. In this context, it becomes crucial to understand how deviations between the model and the real environment impact the optimal cost function and the performance of the perturbed policy, associated with \emph{perturbed transition probability matrices}, which means \emph{perturbed} \mdp s. Perturbation analysis in \rl~examines how changes in the environment or model parameters impact policy performance and value functions. In \cite{geist2010kalman}, a \td~Bayesian approach for value function approximation that models the uncertainty of predictions was proposed, providing robustness against model perturbations.
%In related work reported in \cite{nilim2003robustness} uncertainty sets incorporated for transition probabilities, providing performance guarantees even under worst-case perturbations.
\color{black} 

\color{black}
This paper proves that sufficiently small perturbation in the environment (transition probability matrices) will lead to restricted bound between the optimal cost-to-go function and non-optimal one\color{black}. In particular, we examine how transition probability matrix perturbations can be account as off-policy \td~learning, specifically \texttt{Q}-learning \cite{sutton2018reinforcement}, by investigating the impact of weighted combinations of exploration and optimal actions on efficient learning \cite{sutton2018reinforcement}. In addition, we analyze the variations of cost functions in the context of policy perturbations and provide the corresponding upper bounds for discounted \mdp s\color{black}.
More precisely, let the transition matrices $\mathcal{P}^*$ and $\mathcal{Q}$ \color{black} correspond to the optimal policy $\pi^*$ and an exploratory policy $\pi_{\mathcal{Q}\color{black}}$, respectively. 
Also, let the perturbed transition matrix $\bar{\mathcal{P}}$ be defined as a weighted combination of these two matrices, i.e., for $\bar{\mathcal{P}}$, we have 
\begin{equation}\label{perturb}
\bar{\mathcal{P}} = (I - \mathcal{A}) \mathcal{Q} \color{black}+ \mathcal{A} \mathcal{P}^*,
\end{equation}
where $I$ denotes the identity matrix and $\mathcal{A}$ is a diagonal matrix with $\alpha \in ]0, 1[$ on its diagonal, i.e. $\mathcal{A}=\alpha I$ and $\alpha$ is the discounted factor\color{black}, specifying the trade-off between the use of the optimal policy $\pi^*$ and the exploratory policy $\pi_{\mathcal{Q}\color{black}}$. \color{black} \(\mathcal{A} = \alpha I\) represents a uniform trade-off between the exploratory policy \(\pi_{\mathcal{Q}}\) and the optimal policy \(\pi^*\), but a more general choice of \(\mathcal{A}\) can enable state-dependent perturbations. For instance, by allowing \(\mathcal{A}\) to have different diagonal entries for each state, perturbations can be adjusted adaptively, reflecting varying levels of confidence in the estimated transition dynamics or incorporating state-dependent exploration-exploitation trade-offs. This formulation also aligns well with frameworks that introduce state-dependent learning rates or adaptive exploration strategies, making it particularly useful in reinforcement learning settings where uniform perturbation assumptions may not hold.\color{black}

Our framework explicitly links the perturbation of transition matrices to the exploration-exploitation trade-off through a convex combination of \(\mathcal{Q}\) and \(\mathcal{P}^*\). While similar trade-offs exist in algorithms like \texttt{PPO} and \texttt{TRPO} \cite{schulman2017proximal, schulman2015trust}, the novelty lies in our matrix-based formulation and the explicit derivation of convergence guarantees under perturbations, which is less explored in \mdp~research.
\color{black} We characterize the performance resulting from the utilization of suboptimal policies associated with $\bar{\mathcal{P}}$. 
In particular, we discuss the projection on the subspace of specifically selected features based on such perturbations and the convergence properties for off-policy \td~~methods. 
Furthermore, following the work of \cite{Forootani_IET}, we ensure that the perturbed cost function remains bounded over an infinite time horizon, provided that for any pairs of stationary policy (including $\pi=\pi^*$) it is $\|\mathcal{P} - \bar{\mathcal{P}}\|_\infty \leq 1-\alpha$.

%%%%%%%%%%%%%%%%%%%%%%%%%%%%%%%%%%%%%%%%%%%%%%%%%%%%%%%%%%%%%%

\color{black}
Non-stationary \mdp s have been extensively studied in the \rl~literature, particularly in settings where transition dynamics or reward functions evolve over time. Prior works such as \cite{JMLRjaksch, gajanesliding, cheunghedging, cheungnon, cheungreinforcement, maonearoptimal, ortnervariational, domingueskernel, zhounonstationary, touatiefficient, weinon} focus on different aspects of non-stationarity, including sliding-window-based estimations \cite{cheunghedging, cheungreinforcement, zhongoptimistic}, variation budget constraints \cite{zhounonstationary, weinon}, and kernel-based approaches \cite{domingueskernel}. Most of these works, however, primarily address settings where the transition kernel evolves in an arbitrary or adversarial manner, often requiring explicit tracking mechanisms for adapting to changes. In contrast, our formulation introduces a structured perturbation model where the perturbed transition matrix is represented as a convex combination of the optimal and exploratory transition matrices, as given in \eqref{perturb}. This representation provides a principled approach to modeling controlled deviations in transition dynamics, allowing for a smooth interpolation between optimal and exploratory policies, which aligns with exploration-exploitation trade-offs studied in optimistic \rl~methods \cite{jinprovably, ayoubmodel, zhouprovably}. Furthermore, unlike previous works that rely on tabular settings or parametric drift assumptions \cite{feidynamic, zhounonstationary}, our framework generalizes to off-policy learning scenarios by explicitly linking policy perturbations to transition matrix perturbations, offering a novel perspective on policy optimization under controlled non-stationarity.
\color{black}

This article is organized as follows. Section \ref{background} provides some preliminaries on \dyp~and outlines the cost function approximation problem. The on-policy \adp~approach is discussed in Section \ref{on_policy_adp}. Section \ref{main_results} analyzes the perturbed transition probability matrices and their impacts on cost functions. Section \ref{conection} explores the connection between perturbed transition probability matrices and \texttt{Q}-learning. In Section \ref{simulation}, the proposed off-policy \td~approach is assessed using a resource allocation problem. Section \ref{conclus} concludes the paper.
%%%%%%%%%%%%%%%%%%%%%%%%%%%%%%%%%%%%%%%%
%%%%%%%%%%%%%%%%%%%%%%%%%%%%%%%%%%%%%%%%
\section{Preliminaries}\label{background}
Let us consider an infinite-horizon \dyp~problem for an \color{black}\mdp. Such a problem often is solved by employing Bellman's principle of optimality recursively backwards in time \cite{bertsekas2019reinforcement}. 
In this regard, a shorthand notation is introduced for Bellman operator as $\mathcal{F}^*: \mathbb{R}^{|\mathcal{X}|} \to \mathbb{R}^{|\mathcal{X}|}$ for any cost function vector $J \in \mathbb{R}^{|\mathcal{X}|}$ which can be expressed as \footnote{This approach can be applied similarly for value function optimization by replacing $\min$ function by $\max$ function. See the example in Section VI.}:
\begin{equation}\label{dp_map}
(\mathcal{F}^* J)(x) = \min_{u \in \mathcal{U}} \Big[ \mathcal{R}(x) + \alpha \sum_{x' \in \mathcal{X}} \mathcal{P}_{xx'}(u) J(x') \Big],
\end{equation}
where $\mathcal{X}$ is the \mdp~state space with cardinality $|\mathcal{X}|$, with $x$ and $x'$ being its two generic elements. Moreover, $u \in \mathcal{U}$ is a generic element of finite set of control actions $\mathcal{U}$, $\mathcal{R} \in \mathbb{R}^{|\mathcal{X}|}$ the vector of instant cost with element $\mathcal{R}(x)$\footnote{We omit the dependence of $\mathcal{R}$ on a specific control action.}, $\mathcal{P}:\mathcal{X} \times \mathcal{U} \times \mathcal{X}\to [0,\,1]$ the state transition probability matrix, with generic element $\mathcal{P}_{xx'}(u)$, and $\alpha\in]0,1[$ the discount factor\color{black}. In matrix form, \eqref{dp_map} simplifies to: \(\mathcal{F}^* J = \mathcal{R} + \alpha \mathcal{P}^* J.\)
As noted in \cite{bertsekas2023course}, this shorthand notation is common for simplifying complex expressions. Applying the sequence of optimal decision functions $\{\mu^*, \mu^*, \dots\}$, where $\mu^*:\mathcal{X}\to \mathcal{U}$, as time horizon goes to infinity results in \color{black} optimal stationary policy $\pi^*$  with the associate steady state probability distribution $\epsilon^* \in \mathbb{R}^{|\mathcal{X}|}$(we denote by $\epsilon^*_x$ an element of this vector corresponding to the state $x$)\color{black}. The Bellman equation for the optimal cost function satisfies: \(J^* = \mathcal{R} + \alpha \mathcal{P}^* J^*\), where \( J^* \in \mathbb{R}^{|\mathcal{X}|} \) is the vector of optimal costs, and \( \mathcal{P}^* \in \mathbb{R}^{|\mathcal{X}| \times |\mathcal{X}|} \) is the optimal transition probability matrix. Any non-optimal stationary policy is shown by $\pi$ with its associated shorthand notation $\mathcal{F}_\pi$ and associated cost function $J_\pi$ \cite{bertsekas2019reinforcement, Forootani_IFAC}. The cost function \( J^*(x) \) is usually approximated by a parametric model \( \tilde{J}^*(x, r) \), where \( r \in \mathbb{R}^\psi \) is the parameter vector to be optimized, and $\psi \in \mathbb{N}^+$ is a selected number of features. This approximation, \( \tilde{J}^*: \mathcal{X} \times \mathbb{R}^\psi \to \mathbb{R}^{|\mathcal{X}|} \), uses a low-dimensional linear function \( J^* \approx \Phi r^* \), where \( \Phi \in \mathbb{R}^{|\mathcal{X}| \times \psi} \) is a feature matrix with linearly independent columns  which we show by $\phi(x)$ its row corresponding to state $x$\color{black}, and \( r^* \in \mathbb{R}^\psi \) is the parameter vector to compute. Substituting the approximation into the Bellman equation yields the projected equation: \( \Phi r^* = \Pi (\mathcal{R} + \alpha \mathcal{P}^* \Phi r^*) \), where \( \Pi=\Phi(\Phi^\top \Theta^* \Phi)^{-1} \Phi^\top \Theta^*\) is the projection operator, and $\Theta^*$ is the diagonal matrix with $\epsilon^*$ on the diagonal and zero elsewhere \cite{bertsekasmultiagent}. Here, \( \Phi r^* \) approximates \( J^* \), making the equation solvable. This approach relies on three key assumptions for any stationary policy \( \pi \) (including the optimal \( \pi^* \)) and related cost function $J_\pi$: (i) the corresponding finite Markov chain is regular, with the stochastic matrix \( \mathcal{P} \) having a unique steady-state probability distribution \( \epsilon \), with strictly positive elements $\epsilon_{x}$,  i.e.,  $\epsilon_{x} > 0, \forall x \in \mathcal{X}$; (ii) the feature matrix \( \Phi \) is full rank with rank \( \psi \); (iii) the feature matrix \( \Phi \) captures the key characteristics of the cost function \( J_\pi \), allowing \( \Phi r \) to closely approximate it. \color{black}In this context, the dimensionality of the feature space (\(\psi\)) plays a crucial role in determining computational feasibility. 
\color{black}

\section{\adp~ methods based on On-policy framework}\label{on_policy_adp}
Before delving into the main results of this article, we present introductory material on the on-policy \adp approach to prepare for the discussion of off-policy methods.
In a more general framework, \td~methods can be classified into on-policy and off-policy approaches \cite{sutton2018reinforcement}. In on-policy learning, the goal is to learn the cost function \( \tilde{J}_\pi(x) \), which represents the approximated expected long-term discounted costs from state \( x \) when following a target policy \( \pi \). The learning process occurs while the agent actively follows the same policy \( \pi \). On-policy methods are particularly effective in ensuring good convergence, especially when using linear function approximations. To find the optimal parameter vector \( r^* \) in the on-policy case for the optimal policy $\pi^*$, we solve the following optimization problem:
\begin{equation}\label{adp_infinite}
r^* = \arg \min_{r} \|\Phi r - (\mathcal{R} + \alpha \mathcal{P}^* \Phi r) \|^2_{\epsilon^*},
\end{equation}
which minimizes the square of the weighted Euclidean norm\footnote{ Weighted Euclidean norm on \(\mathbb{R}^{|\mathcal{X}|}\) for any vector $J \in \mathbb{R}^{|\mathcal{X}|}$ is defined as \(\|J\|_{\epsilon^*}=\sqrt{\sum_{\forall x \in \mathcal{X}} \epsilon^*_x \big(J(x)\big)^2}\).\color{black}} \footnote{In this article, we make use of standard definitions of norms for matrices and vectors, e.g., $\|\mathcal{P}\|_{\infty}$ or $\|J\|_{\infty}$.} of the error between \( \Phi r \) and the projected Bellman equation\color{black}, where the weight \( \epsilon^* \) is the steady-state probability distribution \color{black}associated to $\pi^*$\color{black}. By differentiating \eqref{adp_infinite} and setting the gradient to zero, we obtain: 
\( \Phi^\top \Theta^* (\Phi r^* - \mathcal{R} - \alpha \mathcal{P}^* \Phi r^*) = 0, \)
where \( \Theta^* \) is a diagonal matrix with distribution \( \epsilon^*\) along its diagonal. Solving this equation gives:
\begin{align*}\label{linear_system_infinite}
	r^* &= \left(\Phi^\top \Theta^* \Phi - \alpha \Phi^\top \Theta^* \mathcal{P}^* \Phi \right)^{-1} \left(\Phi^\top \Theta^* \right) \mathcal{R}.
\end{align*}
In this formulation, the term \( \left(\Phi^T \Theta^* \Phi - \alpha \Phi^\top \Theta^* \mathcal{P}^* \Phi \right) \) represents the matrix involved in solving for \( r^* \), incorporating both the feature matrix and the transition dynamics. This approach provides a way to approximate the optimal cost function in infinite-horizon problems using a lower-dimensional representation, leveraging the structure of the Bellman equation and the projection framework to find an effective solution. 
Defining $\mathcal{Z}^* = \Phi^\top \Theta^* (I - \alpha \mathcal{P}^*) \Phi$, and $d^* = \Phi^\top \Theta^* \mathcal{R}$, then, the optimal parameter vector is:
\(r^* = \mathcal{Z}^{*-1} d^*\). For large state spaces, computing \( r^* \) directly is infeasible, so iterative methods are employed\color{black}. The update rule can be derived as follows:
\begin{equation*}
	r^*_{k+1} = \arg \min_{r} \|\Phi r - (\mathcal{R} + \alpha \mathcal{P}^* \Phi r_k) \|^2_{\epsilon^*}.
\end{equation*}
Taking the gradient and setting it to zero:
\begin{align*}
	%\Phi^\top \Theta (\Phi r^*_{k+1} - \mathcal{R} - \alpha \mathcal{P}^* \Phi r^*_k) &= 0, \\
	(\Phi^\top \Theta^* \Phi) r^*_{k+1} &= \Phi^\top \Theta^* (\mathcal{R} + \alpha \mathcal{P}^* \Phi r^*_k),
\end{align*}
and rearranging and simplifying:
\begin{equation}\label{on_policy_iter}
	r^*_{k+1} = r^*_k - (\Phi^\top \Theta^* \Phi)^{-1} (\mathcal{Z}^* r^*_k - d^*).
\end{equation}
This iterative update formula is related to Least Squares Temporal Difference (\lstd) methods \cite{bertsekas2023course}. Note that \eqref{on_policy_iter} is applicable to any generic target policy $\pi$ when considering the associated steady state probability distribution \color{black}$\epsilon$ (whose elements will be the diagonal elements of a diagonal matrix $\Theta$). In off-policy learning, the objective remains the same: to estimate the cost function \( J_\pi(x) \) for the target policy \( \pi \). However, the actions taken during the learning process follow a different behavior policy \( \bar{\pi} \). Even though the agent follows \( \bar{\pi} \) during learning, the focus is still on accurately estimating the cost function for the target policy \( \pi \) (see \cite{Sutton_emphatic_2016} and reference therein). The mechanism to compute iteratively the parameter vector is analogous to the one in \eqref{on_policy_iter} and is based on perturbation analysis, which is discussed in the next section.

%%%%%%%%%%%%%%%%%%%%%%%%%%%%%%%
%%%%%%%%%%%%%%%%%%%%%%%%%%%%%%%
%%%%%%%%%%%%%%%%%%%%%%%%%%%%%%%
\section{Perturbation Analysis of stochastic matrices and Cost function bounds}\label{main_results}
In order to apply off-policy methods, which require perturbation of the transition probability matrices (for exploration purposes), it is important to understand how perturbations in transition dynamics affect cost functions. This section establishes bounds for the differences between cost functions from perturbed and unperturbed dynamics, using properties of stochastic matrices and positive definiteness. These results provide insights into the resilience of \mdp s under uncertain transition probabilities, aiding \adp~methods in developing robust decision strategies. \color{black}Using the results reported in \cite{bertsekas2011dynamic} (Lemma 6.3.1)\color{black}, the following Lemma \ref{c_pasitive} and Remark \ref{c_positive_col} guarantee the convergence of on-policy \td~approaches and accordingly recursive iteration \eqref{on_policy_iter}. This result  will be extended to show the convergence property in the case of off-policy \td~approach as well.

\begin{lemma}\label{c_pasitive}
For any stochastic matrix $\mathcal{P}$ corresponding to irreducible and regular Markov chain with associated stationary probability distribution \color{black}$\epsilon$, whose elements are arranged along the diagonal of a diagonal matrix $\Theta$,
the matrix \( \Theta (I - \alpha \mathcal{P})\) is positive definite, where $\alpha \in ]0,1[$.
\end{lemma}

\begin{proof}
To prove that \(\Theta (I - \alpha \mathcal{P})\) is positive definite, we want to show that for any non-zero vector \(J \in \mathbb{R}^{|\mathcal{X}|}\): \(
J^\top \Theta (I - \alpha \mathcal{P}) J > 0.\) We have: 
\(
J^\top \Theta J = \sum_{x} \epsilon_x J_x^2 
\)
and: 
\(
J^\top \Theta \mathcal{P} J =\sum_{x} \epsilon_x J_x \sum_{x'} \mathcal{P}_{xx'} J_{x'},
\)
%\(
%J^\top \Theta \mathcal{P} J = \sum_{x} \epsilon_x J_x \sum_{x'} \mathcal{P}_{xx'} J_{x'}.
%\) To bound this term, let us rewrite it and apply Cauchy-Schwarz inequality. Starting from:
%\[
%\sum_{x} \epsilon_x J_x \sum_{x'} \mathcal{P}_{xx'} J_{x'},
%\]
where the term \(\sum_{x'} \mathcal{P}_{xx'} J_{x'}\) is the expected value of \(J_{x'}\) given \(x\), denoted by \(\mathbb{E}[J_{x'} \mid x]\). Since \(\epsilon\) is a stationary distribution we can rewrite:
\(
\sum_{x} \epsilon_x J_x \sum_{x'} \mathcal{P}_{xx'} J_{x'} = \mathbb{E}_{\epsilon} \left[ J_x \cdot \mathbb{E}[J_{x'} \mid x] \right].\)
Applying the Cauchy-Schwarz inequality to the expectation \(\mathbb{E}_{\epsilon} \left[ J_x \cdot \mathbb{E}[J_{x'} \mid x] \right]\), we get:
\[
\mathbb{E}_{\epsilon} \left[ J_x \cdot \mathbb{E}[J_{x'} \mid x] \right] \le \sqrt{\mathbb{E}_{\epsilon} [J_x^2]} \cdot \sqrt{\mathbb{E}_{\epsilon} \left[(\mathbb{E}[J_{x'} \mid x])^2\right]}.\]

\noindent By Jensen’s inequality for convex functions, we also have:
\[
\mathbb{E}\left[(\mathbb{E}[J_{x'} \mid x])^2\right] \le \mathbb{E}[J_{x'}^2].\] Thus, \[
\sqrt{\mathbb{E}_{\epsilon} \left[(\mathbb{E}[J_{x'} \mid x])^2\right]} \le \sqrt{\mathbb{E}_{\epsilon} [J_x^2]}.\]
Therefore, \(
\mathbb{E}_{\epsilon} \left[ J_x \cdot \mathbb{E}[J_{x'} \mid x] \right] \le \mathbb{E}_{\epsilon} [J_x^2].
\)
Or, equivalently,
\(
\sum_{x} \epsilon_x J_x \sum_{x'} \mathcal{P}_{xx'} J_{x'} \le \sum_{x} \epsilon_x J_x^2.
\) Being $\alpha \in ]0,1[$, this gives us the desired result:
\(
\sum_{x} \epsilon_x J_x^2 >\color{black} \alpha\sum_{x} \sum_{x'} \epsilon_x \mathcal{P}_{xx'} J_x J_{x'}.
\)
\end{proof}

%%%%%%%%%%%%%%%%%%%%%%%%%%%%%%%%%%%%%%%

\begin{remark}\label{c_positive_col}
From the results of Lemma \ref{c_pasitive} and full rank assumption of matrix $\Phi$ we know $\mathcal{Z} = \Phi^\top \Theta (I - \alpha \mathcal{P}) \Phi$ is positive definite. Moreover $(\Phi^\top \Theta \Phi)$ is symmetric positive definite, therefore invertible (see Lemma 5.5 in \cite{forootani2022transmission}). From Proposition 6.3.3 and Lemma 6.3.2 in \cite{bertsekas2011dynamic} the  recursive iteration \eqref{on_policy_iter} directly converges to the solution of the projected equation since the matrix $I - (\Phi^\top \Theta \Phi)^{-1} \mathcal{Z}$ has eigenvalues strictly within the unit \color{black}circle (see also Theorem 5.6 in \cite{forootani2022transmission}).
\end{remark}

Remark \ref{c_positive_col} implies that necessary and sufficient condition for the convergence of any on-policy \td~algorithm depends on the positive definiteness of matrix $\mathcal{Z}$, and hence of $\Theta (I-\alpha \mathcal{P})$. This concept can be extended to the case of off-policy \td~algorithm. In the next Lemma, we investigate such property for the case of perturbed transition probability matrix $\bar{\mathcal{P}}$.

%%%%%%%%%%%%%%%%%%%%%%%%%%%%%%%%%%%%%%

\begin{lemma}\label{c_purturbed}
For any stochastic matrix \(\mathcal{P}\) corresponding to an irreducible and regular Markov chain, consider a stationary probability distribution \color{black}\(\bar{\epsilon}\), whose elements are arranged along the diagonal of a diagonal matrix $\bar{\Theta}$, associated with the perturbed matrix \(
\bar{\mathcal{P}} =  (I - \mathcal{A}) \mathcal{Q} \color{black} +\mathcal{A} \mathcal{P},
\) being \(\mathcal{A}\) a diagonal matrix with \(\alpha\in]0,1[\) on its diagonal, and \(\mathcal{Q}\color{black} \) another stochastic matrix. Then, the matrix \( \bar{\Theta} (I - \alpha \mathcal{P})
\) is positive definite.
\end{lemma}\label{positive_definite_bar} 
\begin{proof}
We need to show that for any non-zero vector \(J \in \mathbb{R}^{|\mathcal{X}|}\), \( J^\top \bar{\Theta} (I - \alpha \mathcal{P}) J > 0 \). To this end, consider the matrix  $\bar{\Theta} (I-\alpha \bar{\mathcal{P}})$, and %Jensen’s inequality for convex functions applied to \(\bar{\mathcal{P}}\), since \(\bar{\mathcal{P}}_{xx'} \geq 0\) and \(\sum_{x'} \bar{\mathcal{P}}_{xx'} = 1\), then \( \sum_{x'} \bar{\mathcal{P}}_{xx'} J_{x'}^2 \ge \big( \sum_{x'} \bar{\mathcal{P}}_{xx'} J_{x'} \big)^2.\)
%Therefore (from
apply the results of Lemma \ref{c_positive_col}. Following the same proof steps,  we can state that:
\begin{equation}\label{conv_bar_p}
\sum_{x} \bar{\epsilon}_x J_x^2 \ge \sum_{x,x'} \bar{\epsilon}_x \bar{\mathcal{P}}_{xx'} J_x J_{x'},
\end{equation}
and also 
\(
\sum_x \bar{\epsilon}_x J_x^2 - \alpha \sum_x \bar{\epsilon}_x \sum_{x'} \frac{\bar{\mathcal{P}}_{xx'}}{\alpha} J_x J_{x'} \ge 0 \). 
Since $\bar{\mathcal{P}}_{xx'}=\alpha \mathcal{P}_{xx'}+(1-\alpha)\mathcal{Q}\color{black}_{xx'}$, and being $\mathcal{Q}\color{black}$ another stochastic matrix, by substitution in \eqref{conv_bar_p}\color{black}, it is:
\( \sum_x \bar{\epsilon}_x J_x^2 - \alpha \sum_x \bar{\epsilon}_x \sum_{x'} \mathcal{P}_{xx'} J_x J_{x'} > 0,
\)
which completes the proof.
\end{proof}

Now we are ready to consider the impact of perturbation introduced by the matrix \(\bar{\mathcal{P}}\), defined as \(\bar{\mathcal{P}} = (I - \mathcal{A}) \mathcal{Q} \color{black}+ \mathcal{A} \mathcal{P}^*\), on the \color{black}projected Bellman equation. This perturbation can be incorporated into the projected Bellman equation \color{black}by considering the perturbed transition matrix \(\bar{\mathcal{P}}\) instead of \(\mathcal{P}^*\). The projected Bellman equation with perturbation is:
\( \Phi \bar{r} = \bar{\Pi} \left(\mathcal{R} + \alpha \mathcal{P}^* \Phi \bar{r}\right),\)
where \(\bar{\Pi} = \Phi (\Phi^\top \bar{\Theta} \Phi)^{-1} \Phi^\top \bar{\Theta}\) is the projection operator. Substituting \(\bar{\Pi}\) and rearranging, we get:	
\( \left(\Phi^\top \bar{\Theta} \Phi - \alpha \Phi^\top \bar{\Theta} \mathcal{P}^* \Phi\right) \bar{r} = \Phi^\top \bar{\Theta} \mathcal{R}.\) Defining \( \bar{\mathcal{Z}} = \Phi^\top \bar{\Theta} \left(I - \alpha \mathcal{P}^* \right) \Phi,
	\bar{d} = \Phi^\top \bar{\Theta} \mathcal{R},
\)
so the solution for \(\bar{r} \) is: \( \bar{r} = \bar{\mathcal{Z}}^{-1} \bar{d}.\)  
This formulation allows us to solve for the parameter vector \(\bar{r}\) that approximates the cost function under the perturbed policy, providing a way to assess the impact of deviations from the optimal policy within the framework of \adp. Similar to previous case, for large state spaces, solving directly for \( \bar{r} \) is impractical. Instead, iterative methods like \td~are used. The iterative update formula in the off-policy case is:
\begin{equation}\label{recursive_r}
	\bar{r}_{k+1} = \bar{r}_k - (\Phi^\top \bar{\Theta} \Phi)^{-1} ( \bar{\mathcal{Z}} r_k - \bar{d} ),
\end{equation}
whose convergence property can be derived from the following theorem. 

%%%%%%%%%%%%%%%%%%%%%%%%%%%%%%%%%%%%%%%%%%%%
%%%%%%%%%%%%%%%%%%%%%%%%%%%%%%%%%%%%%%%%%%%%

\begin{comment}
\begin{theorem}
Consider an \mdp~with $\mathcal{X},\ \mathcal{U},\ \mathcal{R}$ and optimal transition probability matrix $\mathcal{P}^*$. Let $\bar{\mathcal{P}}=(I-\mathcal{A})\mathcal{Q}\color{black}+\mathcal{A} \mathcal{P}^*$ be the transition probability matrix associated to a perturbed policy. Then the recursive iteration \eqref{recursive_r} converges. 
\begin{proof}
From the results of Lemma  \ref{c_purturbed} for the case $\mathcal{P}=\mathcal{P}^*$, we know $\bar{\mathcal{Z}} = \Phi^\top \bar{\Theta} (I - \alpha \mathcal{P}^*) \Phi$ is positive definite. Moreover $(\Phi^\top \bar{\Theta} \Phi)$ is symmetric positive definite, therefore invertible. Since the convergence of \eqref{recursive_r} is directly related to the positive definiteness of the term $\bar{\Theta} (I - \alpha \mathcal{P}^*)$, then we can guarantee the convergence of projected based \adp~with perturbed transition matrices. 
\end{proof}
\end{theorem}
\end{comment}

\begin{theorem}\label{off_policy_theorem}
	Consider a Markov Decision Process (\mdp) with state space \(\mathcal{X}\), action space \(\mathcal{U}\), reward vector \(\mathcal{R}\), and optimal transition probability matrix \(\mathcal{P}^*\). Let the perturbed transition probability matrix be defined as:
	\(
	\bar{\mathcal{P}} = (I - \mathcal{A})\mathcal{Q} + \mathcal{A} \mathcal{P}^*,
	\)
	where \(\mathcal{A}\) is a diagonal matrix with the discount factor \(\alpha \in ]0, 1[\) along its diagonal, and \(\mathcal{Q}\) is a stochastic matrix associated with an exploratory policy. Define:
	\(
	\bar{\mathcal{Z}} = \Phi^\top \bar{\Theta} \left(I - \alpha \mathcal{P}^*\right) \Phi, \quad \bar{d} = \Phi^\top \bar{\Theta} \mathcal{R},
	\)
	where \(\Phi \in \mathbb{R}^{|\mathcal{X}| \times m}\) is a feature matrix, and \(\bar{\Theta}\) is a diagonal matrix with the stationary probability distribution \(\bar{\epsilon}\) on its diagonal. The recursive iteration for the parameter vector \(\bar{r}_k \in \mathbb{R}^\psi\) is given by:
	\(
	\bar{r}_{k+1} = \bar{r}_k - (\Phi^\top \bar{\Theta} \Phi)^{-1} \left( \bar{\mathcal{Z}} \bar{r}_k - \bar{d} \right).
	\)
	Then, the iterative update \(\bar{r}_k\) converges to the solution \(\bar{r}\) of the projected Bellman equation:
	\(
	\Phi \bar{r} = \bar{\Pi} \left(\mathcal{R} + \alpha \mathcal{P}^* \Phi \bar{r}\right),
	\) where \(\bar{\Pi} = \Phi (\Phi^\top \bar{\Theta} \Phi)^{-1} \Phi^\top \bar{\Theta}\).
\begin{proof}
From the results of Lemma  \ref{c_purturbed} for the case $\mathcal{P}=\mathcal{P}^*$, we know $\bar{\mathcal{Z}} = \Phi^\top \bar{\Theta} (I - \alpha \mathcal{P}^*) \Phi$ is positive definite. Moreover $(\Phi^\top \bar{\Theta} \Phi)$ is symmetric positive definite, therefore invertible. Since the convergence of \eqref{recursive_r} is directly related to the positive definiteness of the term $\bar{\Theta} (I - \alpha \mathcal{P}^*)$, then we can guarantee the convergence of projected based \adp~with perturbed transition matrices. 
\end{proof}
\end{theorem}
\color{black}

%%%%%%%%%%%%%%%%%%%%%%%%%%%%%%%%%%%%%%%%%%%%
%%%%%%%%%%%%%%%%%%%%%%%%%%%%%%%%%%%%%%%%%%%%

\begin{remark}\label{monte_carlo}
The recursive iteration \eqref{recursive_r} can be computed using the Monte Carlo simulation mechanism. This procedure involves producing two sequence of state visits via off-policy \td~methods, where the sequence \( \big\{ \bar{x}(0), \bar{x}(1), \bar{x}(2), \ldots \big\} \) is generated using the transition matrix \( \bar{\mathcal{P}} \) (behavior policy) or a steady-state distribution \( \bar{\epsilon}\), and we also generate an additional sequence of independent transitions \( \big\{ \big( \bar{x}(0), x(0) \big), \big( \bar{x}(1) , x(1) \big), \ldots \big\} \) according to an original transition matrix \( \mathcal{P} \) not necessarily optimal (target policy). %The off-policy \td~approach that we employ in this paper is based on the method in \cite{forootani2022transmission} namely Enhanced-exploration Greedy \td, with mainly difference in the selection criteria associated to behavior policy as proposed in this article.
\end{remark}

%%%%%%%%%%%%%%%%%%%%%%%%%%%%%%%%%%%%%%%%%%%%%
%%%%%%%%%%%%%%%%%%%%%%%%%%%%%%%%%%%%%%%%%%%%%
%%%%%%%%%%%%%%%%%%%%%%%%%%%%%%%%%%%%%%%%%%%%%
An important metric in analyzing perturbed \mdp s is the upper bound for the difference between cost functions derived from perturbed and unperturbed transition probabilities matrices. This ensures consistent policy performance under non-stationary or perturbed conditions in \mdp s and helps manage approximation errors in value iteration and \adp~methods, thereby maintaining optimal policy quality \cite{schulman2017proximal}.

The next theorem provides an upper bound for the difference between optimal cost function and its perturbed counterpart. This result is particularly useful when the optimal policy is not known and we want to obtain a bound on the norm of the error between the estimated cost function and optimal one.

\begin{theorem}\label{TheoMain}
Consider an \mdp \, with  $\mathcal{X}, \mathcal{U}, \mathcal{R}$,  and  two different transition probability matrices,  \(\mathcal{P}^*\) and \(\bar{\mathcal{P}}\), with associated cost functions \(J^*\) and \(\bar{J}\), respectively. Let \(\bar{\mathcal{P}} = (I - \mathcal{A}) \mathcal{Q}\color{black} + \mathcal{A} \mathcal{P}^*\), where \(\mathcal{A}\) is a diagonal matrix with \(\alpha\in]0,1[\) on the diagonal, and \(\mathcal{Q}\color{black}\) is a stochastic matrix. If \(\|\mathcal{P}^* - \bar{\mathcal{P}}\|_\infty \le (1 - \alpha)\), the difference between \(J^*\) and \(\bar{J}\) is bounded by:
\(
\|J^* - \bar{J}\|_\infty \leq \frac{\alpha \|\mathcal{R}\|_\infty}{1 - \alpha}.
\)
\end{theorem}

\begin{proof}
Using the Bellman optimality equations and the fixed point mapping of the \dyp~for \(J^*\) and \(\bar{J}\) \cite{bertsekas2019reinforcement}, we can write the following: 
%\begin{multline*}
%\|J^*(x) - \bar{J}(x)\| = \big\|  \big[ \mathcal{R}(x) + \alpha \sum_{x' \in \mathcal{X}} \mathcal{P}^*(x' | x, u) J^*(x') \big] \\ -  \big[ \mathcal{R}(x) + \alpha \sum_{x' \in \mathcal{X}} \bar{\mathcal{P}}(x' | x, u) \bar{J}(x') \big] \big\|\color{black}.
%\end{multline*}
%Using the properties of the \(\min\) function (Proposition \ref{minima_prop}), we can bound this difference
%by selecting a common state \(x\) for both value/cost functions as follows:
\begin{equation}
 |J^*(x) - \bar{J}(x)| = \big | \mathcal{R}(x) + \alpha \sum_{x' \in \mathcal{X}} \mathcal{P}^*_{xx'} J^*(x') \\- \mathcal{R}(x) - \alpha \sum_{x' \in \mathcal{X}} \bar{\mathcal{P}}_{xx'} \bar{J}(x') \big|.   
\end{equation}

\color{black}
This simplifies to:
\begin{equation}\label{first_step}
\big |J^*(x) - \bar{J}(x)\big| = \alpha \big | \sum_{x' \in \mathcal{X}} \mathcal{P}^*_{xx'} J^*(x')- \sum_{x' \in \mathcal{X}} \bar{\mathcal{P}}_{xx'} \bar{J}(x') \big |.
\end{equation}\color{black}
We now decompose the sum:
\begin{equation}\label{mediate_step}
\sum_{x' \in \mathcal{X}} \mathcal{P}^*_{xx'} J^*(x') - \sum_{x' \in \mathcal{X}} \bar{\mathcal{P}}_{xx'} \bar{J}(x) = \sum_{x' \in \mathcal{X} } \big( \mathcal{P}^*_{xx'} J^*(x') \\- \bar{\mathcal{P}}_{xx'} J^*(x') \big) + \sum_{x' \in \mathcal{X}} \bar{\mathcal{P}}_{xx'} (J^*(x') - \bar{J}(x')).
\end{equation}

Substituting \eqref{mediate_step} into \eqref{first_step} we have:
\begin{equation*}
\big|J^*(x) - \bar{J}(x)\big| = \alpha \big|\sum_{x' \in \mathcal{X} } \left( \mathcal{P}^*_{xx'} J^*(x') - \bar{\mathcal{P}}_{xx'} J^*(x') \right) \\ + \sum_{x' \in \mathcal{X}} \bar{\mathcal{P}}_{xx'} (J^*(x') - \bar{J}(x'))\big|,
\end{equation*}
Using \color{black} triangle \color{black} inequality we get:
\color{black}
\begin{equation}\label{traiangular_ineq}
|J^*(x) - \bar{J}(x)| \leq \alpha \sum_{x' \in \mathcal{X}} |\mathcal{P}^*_{xx'} - \bar{\mathcal{P}}_{xx'}| \ |J^*(x')\big| \\ + \alpha \sum_{x' \in \mathcal{X}} |\bar{\mathcal{P}}_{xx'}| \ |J^*(x') - \bar{J}(x')|,
\end{equation}
where \color{black}the first term contains the difference between the transition matrices \(\mathcal{P}^*\) and \(\bar{\mathcal{P}}\), while the second term accounts for the difference in cost functions.
%:
%\begin{multline}
%\big\|J^*(x) - \bar{J}(x)\big\| \leq \alpha \sum_{x' \in \mathcal{X}} \big|\mathcal{P}^*_{xx'} - \bar{\mathcal{P}}_{xx'}\big|\ \big|J^*(x')\big| \\ + \alpha \sum_{x' \in \mathcal{X}} \big|\bar{\mathcal{P}}_{xx'}\big| \ \big|J^*(x') - \bar{J}(x')\big|.
%\end{multline}
Using the assumption that \(\big\|\mathcal{P}^* - \bar{\mathcal{P}}\big\|_\infty < (1 - \alpha)\), the properties of infinite norm, \color{black} and the fact that \eqref{traiangular_ineq} holds for any norm, we bound the first term:
\begin{equation*}
\|J^* - \bar{J} \|_\infty \leq \alpha (1 - \alpha) \|J^*\|_\infty + \alpha \sum_{x' \in \mathcal{X}} |\bar{\mathcal{P}}_{xx'}| |J^*(x') - \bar{J}(x')|.
\end{equation*}
Next, we use the fact that \(\sum_{x' \in \mathcal{X}} \bar{\mathcal{P}}_{xx'} = 1\), since \(\bar{\mathcal{P}}\) is a stochastic matrix. Therefore, we have:
\[
\big\|J^* - \bar{J} \big\|_{\infty} \color{black} \leq \alpha (1 - \alpha) \big\|J^*\big\|_\infty + \alpha \big\|J^* - \bar{J}\big\|_\infty,
\]
Rearranging the inequality to isolate \(\|J^* - \bar{J}\|_\infty\), we get:
\(
\big\|J^* - \bar{J}\big\|_\infty \leq \frac{\alpha (1 - \alpha) \big\|J^*\big\|_\infty}{1 - \alpha} 
\).
Since we know from the Bellman equation that \(\|J^*\|_\infty \leq \frac{\|\mathcal{R}\|_\infty}{1 - \alpha}\), by substituting this bound, we obtain:
\(
\|J^* - \bar{J}\|_\infty \leq \frac{\alpha \|\mathcal{R}\|_\infty}{1 - \alpha},
\)
which completes the proof.
\end{proof}
\begin{remark}
The norm condition $\|\mathcal{P}^* - \bar{\mathcal{P}}\|_\infty \le (1 - \alpha)$ in Theorem \ref{TheoMain} is satisfied whenever it is $ \|\mathcal{P}^*-\mathcal{Q}\color{black}\|_\infty \le 1$. Indeed, the deviation from the optimal policy introduces a perturbation matrix 
\(\mathcal{P}^*-\bar{\mathcal{P}} = (I - \mathcal{A}) (\mathcal{P}^* - \mathcal{Q}\color{black}),\) whose norm is bounded as follows:  \(\|\mathcal{P}^* - \bar{\mathcal{P}}\|_\infty \le (1 - \alpha) \|\mathcal{P}^*-\mathcal{Q}\color{black}\|_\infty\).     
\end{remark}

%%%%%%%%%%%%%%%%%%%%%%%%%%%%%%%%%%%%%%%%%%%%
%%%%%%%%%%%%%%%%%%%%%%%%%%%%%%%%%%%%%%%%%%%%
%%%%%%%%%%%%%%%%%%%%%%%%%%%%%%%%%%%%%%%%%%%%
\color{black}

\begin{remark}
While Theorems \ref{off_policy_theorem} and \ref{TheoMain} analyze the impact of perturbations on the optimal transition matrix $\mathcal{P}^*$, the proposed algorithm is applicable to any general target transition matrix $\mathcal{P}$, corresponding to an arbitrary stationary policy, whether optimal or suboptimal. Since off-policy \td~learning is independent of whether $\mathcal{P}$ is optimal, our framework extends beyond optimal policies and remains valid for exploratory and intermediate policies as well.
\end{remark}
\color{black}

Algorithm \ref{alg:off_policy_td} summarizes the off-policy implementation using \td~learning and perturbed transition matrix $\bar{\mathcal{P}}$ for the general target transition probability matrix $\mathcal{P}$.
%\color{green} Considering Remark 2, we should have in the step 1 of Algo1 the target policy $\mathcal{P}$ as input, and then $\mathcal{P}$ in step 5.\color{black}
\begin{algorithm}
\small
\caption{Off-Policy \td~Learning with $\bar{\mathcal{P}}$}
\label{alg:off_policy_td}
\begin{algorithmic}[1]
\STATE Initialize parameter vector $\bar{r}_0$, discount factor $\alpha$, feature representation $\phi(x)$ for each state $x$, and behavior policy transition matrix $\bar{\mathcal{P}}$
\FOR{each episode}
    %\STATE Generate an initial state $\bar{x}(0)$ using behavior policy or steady-state distribution $\bar{\epsilon}$
    %\STATE Generate an initial state $x(0)$ via an independent target policy transition
    \FOR{each time step $k$}
        \STATE Generate the state $\bar{x}(k)$ using $\bar{\mathcal{P}}$
        \STATE Generate the state $x(k)$ from the transition pair $\big(\bar{x}(k), x(k)\big)$ using $\mathcal{P}$
        \STATE Observe the immediate cost $\mathcal{R}(x(k))$ associated with the target policy
        \STATE Compute the \td~error:
        \[
        \delta_k = \mathcal{R}(x(k)) + \alpha \phi(x(k))^\top \bar{r} - \phi(\bar{x}(k))^\top \bar{r}
        \]
        \STATE Update parameter vector: \(
        \bar{r} \leftarrow \bar{r} + \delta_k \phi(x_k)
        \)
        %[Check if it's $r$ or $\bar{r}$]\color{black}
        %\STATE Ensure the independence of $\big(\bar{x}(k), x(k)\big)$ in the transition sequences
    \ENDFOR
\ENDFOR
\end{algorithmic}
\end{algorithm}
\normalsize

%%%%%%%%%%%%%%%%%%%%%%%%%%%%%%%%%%%%%
%%%%%%%%%%%%%%%%%%%%%%%%%%%%%%%%%%%%%

\section{Analogy Between the Perturbed Transition Matrix and \texttt{Q}-Learning}\label{conection}
In this section, we discuss how the analysis on 
perturbed transition matrix \(\bar{\mathcal{P}}\) relates to \texttt{Q}-learning, where the optimal policy \(\pi^*\) is combined with an exploratory policy, say \(\pi_{\mathcal{Q}\color{black}}\). The mixture of these two policies can be expressed as a weighted combination, similar to the exploration-exploitation trade-off in \texttt{Q}-learning. In \texttt{Q}-learning, the policy is a mixture of an \textit{exploratory policy} and an \textit{optimal policy}, in the sense that the action-selection strategy in Q-learning combines exploitation of the optimal policy \(\pi^*\) and exploration using a stochastic policy \(\pi_{\mathcal{Q}\color{black}}\). 

This is mathematically described by the policy mixture \cite{sutton2018reinforcement}: \[\pi_{\xi}(u|x) = \xi \cdot \pi_{\mathcal{Q}\color{black}}(u|x) + (1 - \xi ) \cdot \pi^*(u|x),\] 
where \(\pi^*(u|x)\) is the optimal policy that chooses action \(u\) in state \(x\) based on the optimal \texttt{Q}-function \(\texttt{Q}^*(x, u)\), \(\pi_{\mathcal{Q}\color{black}}(u|x)\) is an exploratory policy, and \(\xi \in ]0, 1[\) is the exploration parameter controlling the balance between exploration and exploitation. Under this mixed policy \(\pi_{\xi}\), the transition dynamics can be expressed as a combination of the transition matrices corresponding to \(\pi^*\) and \(\pi_{\mathcal{Q}\color{black}}: P_{\pi_{\xi}} = \xi P_{\pi_{\mathcal{Q}\color{black}}} + (1 - \xi) P_{\pi^*}.\) We now observe that the perturbed transition matrix \(\bar{\mathcal{P}}\) can be seen as a similar mixture of policies, with \(\alpha\) playing the role of \(1 - \xi\) in \texttt{Q}-learning:
\[
\bar{\mathcal{P}} = (I - \mathcal{A}) \mathcal{Q} \color{black}+ \mathcal{A} \mathcal{P}^*.\]
This formulation shows that \(\bar{\mathcal{P}}\) models the transition dynamics of a perturbed policy \(\bar{\pi}\), which can be described as a probabilistic mixture of the optimal policy \(\pi^*\) and the exploratory policy \(\pi_{\mathcal{Q}\color{black}}\): \( \bar{\pi}(u|x) = (1 - \alpha) \pi_{\mathcal{Q}\color{black}}(u|x) + \alpha \pi^*(u|x)\)\footnote{The policy \(\bar{\pi}\) deviates from \(\pi^*\) to \(\pi_{\mathcal{Q}\color{black}}\) with probability \(1 - \alpha\).}. Thus, \(\alpha\) acts similarly to \(1 - \xi\), determining how much the system follows the optimal policy \(\mathcal{P}^*\) versus exploring the exploratory transitions defined by \(\mathcal{Q}\color{black}\). Since \(\bar{\mathcal{P}}\) is a weighted combination of the optimal transition matrix \(\mathcal{P}^*\) and the exploratory matrix \(\mathcal{Q}\color{black}\), the deviation from the optimal transition matrix is given by: \( \|\mathcal{P}^* - \bar{\mathcal{P}}\|_\infty = (1-\alpha) \|\mathcal{P}^* - \mathcal{Q}\color{black}\|_\infty\)\color{black}.

Assuming that \(\|\mathcal{P}^* - \mathcal{Q}\color{black}\|_\infty \leq 1\), this deviation is bounded by: \(
\|\mathcal{P}^* - \bar{\mathcal{P}}\|_\infty \leq 1-\alpha\). This expression is analogous to the temporal difference error in \texttt{Q}-learning, where \(\alpha\) controls the trade-off between exploration and exploitation \cite{sutton2018reinforcement}. As \(\alpha\) decreases, the deviation from the optimal transition dynamics increases, reflecting a greater degree of exploration.
%The matrix \(\bar{\mathcal{P}} = (1 - \mathcal{A}) Q + \mathcal{A} \mathcal{P}^*\) mirrors the exploration-exploitation trade-off in \texttt{Q}-learning, where \(\alpha\) acts as the counterpart to \(1 - \xi\).
This trade-off allows the system to balance between following the optimal policy \(\pi^*\) and exploring alternative policies through \(\pi_{\mathcal{Q}\color{black}}\).

The deviation from the optimal transition matrix \(\mathcal{P}^*\) is controlled by \(\alpha\), similar to the exploration parameter in \texttt{Q}-learning. While the principles of perturbation in Q-learning and the presented approach share similarities, they are \textit{not interchangeable} in general. Q-learning excels in model-free and discrete setups, while our approach is better suited for structured, large-scale, or robust policy optimization tasks. These differences make them complementary tools in the \rl ~framework.\color{black}

\section{Numerical Simulations}\label{simulation}
In this section, we compute an estimation of the optimal value function $J^*$ through $\bar{J}$, and hence the parameter vector $\bar{r}$, utilizing the off-policy \td~approach discussed earlier for an \mdp~related to a resource allocation problem \cite{Forootani_IFAC, Forootani_cont_sys_lett}\color{black}.
 In summary, the challenge of managing resource pricing for dynamic allocation is tackled by utilizing parallel stochastic Markov chains. In this model, customers request a resource with a probability of \( \lambda_i \) and release it with a probability of \( \mu_i \), both of which correspond to the price \( c_i \). The decision-maker's objective is to maximize profit over an infinite time horizon, which is analogous to minimizing costs in the dual formulation of the problem. \color{black}Denoting by $x_i(k)$ the state of Markov chain (here number of customers) associated to $c_i$, we can compute the value function as follows:
\begin{equation*}
J^*\big(x\big)= \max_{ u(k) \in \mathcal{U}} \lim_{T \to \infty} \mathbb{E}\left[ \sum_{k=0}^{T} \alpha^k\sum_{i=1}^{m}  c_i x_i(k) \right],
\end{equation*}
where $\mathcal{U} = \{c_1, c_2, \ldots, c_m\}$. This resource allocation problem exhibits the curse of dimensionality as the number of available resources $N$ and the possible price choice $m$ increase (see \cite{Forootani_IFAC} for details).

%%%%%%%%%%%%%%%%%%%%%%%%%%%%%%%%%%%%%%%%%%%%%%%%%%%%%
%%%%%%%%%%%%%%%%%%%%%%%%%%%%%%%%%%%%%%%%%%%%%%%%%%%%%

\color{black}
To provide more information and to elaborate more (see \cite{forootani2020least, forootani2021modelling, Forootani_IET}), by exploiting Bellman's principle of optimality, the optimal value function for the tail subproblem at time \(k\) can be written as
\begin{align}\label{dp_optimality}
	J^*_k\bigl(x(k)\bigr) 
	&=  \max_{ u(k) \in \mathcal{U}} \mathbb{E}\left[ \sum_{k=0}^{T} \sum_{i=1}^{m} \alpha^k c_i x_i(k) \right] \nonumber \\ &= \max_{u(k)\in \{c_1,c_2,\dots,c_m\} }
	\mathbb{E}\Bigl[
	\sum_{i=1}^m c_i \, x_i(k) 
	+\alpha J^*_{k+1}\bigl(x(k+1)\bigr)
	\Bigr]. \nonumber
\end{align}

More specifically, by applying Bellman's recursive backward procedure and starting from the terminal condition
\[
J_T\bigl(x(T)\bigr) 
= \sum_{i=1}^{m} c_i \, x_i(T),
\]
the optimal value function for such a tail subproblem can be expressed as
\begin{align*}
	J^*_{k}\bigl(x(k)\bigr)
	&= \max_{ u(k)\in \{c_1,c_2,\dots,c_m \} } 
	\mathbb{E}\Bigl\{\sum_{i=1}^{m} c_i \, x_i(k)
	\\ &+\alpha J^*_{k+1}\bigl(x(k+1)\bigr)\Bigr\} = \sum_{i=1}^{m} c_i \, x_i(k)
	\;+\; \\ & \alpha \max_{u(k)\in \{c_1,c_2,\dots,c_m\} }
	\mathbb{E}\Bigl\{J^*_{k+1}\bigl(x(k+1)\bigr)\Bigr\} \\ &= \sum_{i=1}^{m} c_i \, x_i(k)
	\;+\;\\ & \alpha \max \biggl[
	\mathbb{E} \Bigl\{   J^*_{k+1}\bigl(x(k+1)\bigr) \,\big|\;u(k)=c_1  \Bigr\},
	\;\dots,\;\\
	&\mathbb{E} \Bigl\{   J^*_{k+1}\bigl(x(k+1)\bigr) \,\big|\;u(k)=c_m  \Bigr\}
	\biggr].
\end{align*}

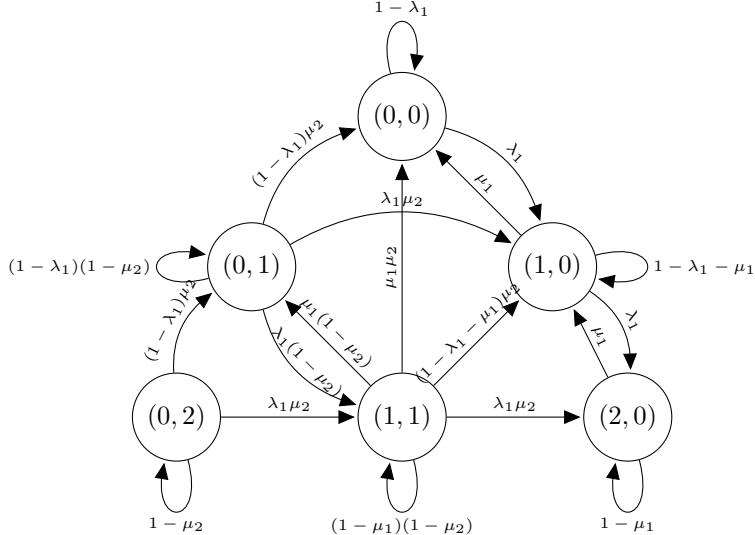
\begin{figure}[h!]
	\color{black}
	\centering
	%\tiny
	\small
	\begin{tikzpicture}[->,>=triangle 45,shorten >=1pt,auto,
	thin]
	%\tikzstyle{every state}=[fill=red,draw=none,text=white]
	%\draw  (-6,6) rectangle (6,-6);
	%\clip(-6,6) rectangle (6,-6);
	%\node[state] 		 (A) [] at (-3,5) {$\xi^0_1$};
	
	\node[state] 		 (A) [] at (-1,0) {$(0,0)$};
	\node[state]         (B) [] at(1,-2) {$(1,0)$};
	\node[state]     	 (C) [] at (2,-4) {$(2,0)$};
	
	\node[state]         (D) [] at (-3,-2) {$(0,1)$};
	\node[state]         (E) [] at (-1,-4) {$(1,1)$};
	\node[state]         (F) [] at (-4,-4) {$(0,2)$};

	\tiny
	\path (A) edge  [bend left]    node [pos=0.5, sloped, above] {$\lambda_1$} (B)
	(B) edge  []    node [pos=0.5, sloped, above] {$\mu_1$} (A)
	(B) edge  [bend left]    node [pos=0.5, sloped, above] {$\lambda_1$} (C)
	(B) edge [loop right] node {$1-\lambda_1-\mu_1$} (B)
	(A) edge [loop above] node {$1-\lambda_1$} (A)
	(C) edge  []   node [pos=0.5, sloped, above] {$\mu_1$} (B)
	%(C) edge  [bend left]    node {$\lambda_1$} (D)
	(C) edge [loop below] node {$1-\mu_1$} (C)
	(D) edge  [bend left] node [pos=0.5, sloped, above] {$(1-\lambda_1)\mu_2$} (A)
	(D) edge  [bend right]    node [pos=0.5, sloped, above]{$\lambda_1(1-\mu_2)$} (E)
	(D) edge [loop left] node {$(1-\lambda_1)(1-\mu_2)$} (D)
	(D) edge [bend left] node {$\lambda_1\mu_2$} (B)
	(E) edge  []    node [pos=0.5, sloped, above] {$(1-\lambda_1-\mu_1)\mu_2$} (B)
	(E) edge [] node [pos=0.5, sloped, above] {$\mu_1 (1-\mu_2)$} (D)
	(E) edge [] node [pos=0.5, sloped, above] {$\mu_1\mu_2$} (A)
	(E) edge  []    node {$\lambda_1\mu_2$} (C)	
	(F) edge [loop below] node {$1-\mu_2$} (F)
	(F) edge  [bend left]    node [pos=0.5, sloped, above] {$(1-\lambda_1)\mu_2$} (D)
	(F) edge  []    node {$\lambda_1\mu_2$} (E)
	(E) edge  [loop below]    node {$(1-\mu_1)(1-\mu_2)$} (E)
	;
	\end{tikzpicture}
	
	\caption{\textcolor{black}{The graph shows the MDP state space and the corresponding state transition probabilities for a resource allocation problem with $N=2$ and $m=2$, and for the control input $u(k)=c_1$. The system states are $(x_1,x_2)\in \{(0,0),(0,1),(1,0),(1,1),(0,2),(2,0)\}$, where $x_1$ and $x_2$ are associated to $c_1$ and $c_2$, respectively.}} 
	\label{MDP_example}
\end{figure}

Finally, \(J_0\bigl(x(0)\bigr)\), generated at the last step, is equal to the optimal value function \(J^*\bigl(x(0)\bigr)\) computed for the entire finite time horizon \(T\). Fig. \ref{MDP_example} shows an example of the \mdp~associated to a resource allocation problem for a given policy.

As illustrated in Table \ref{curse_dimensionality} as the number of resources \(N\) and the number of actions \(m\) increase, the cardinality of the state space grows exponentially. For instance, if \(m=5\) and \(N=20\), performing just one iteration of Bellman's recursion to solve the exact Dynamic Programming (DP) problem entails evaluating \(5 \times 53130\) action-state pairs \((u, x)\) and computing their associated transition probabilities \(\mathcal{P}_{xx'}(u)\), where \(\mathcal{P} \colon \mathcal{X} \times \mathcal{U} \times \mathcal{X}\).

\begin{table}[h]
\color{black}
\centering
\caption{Demonstrating the curse of dimensionality in resource allocation problem}
\label{curse_dimensionality}
	{\begin{tabular*}{20pc}{@{\extracolsep{\fill}}lll@{}}\toprule
			\multicolumn{2}{c}{\textbf{Example of state space explosion [3]}} \\
			\midrule
			$m$ & $N$ & \textit{Number of States}\\
			2 & 10 & 66 \\
		
			3 & 20 & 1\,771\\
			
			5 & 20 & 53\,130\\
		
			6 & 20 & 230\,230\\
			
			4 & 50 & 316\,251\\
			
			5 & 50 & 3\,478\,761\\
			
	\end{tabular*}}{}
    
\end{table}

\color{black}

%%%%%%%%%%%%%%%%%%%%%%%%%%%%%%%%%%%%%%%%%%%%%%%%%%%%
%%%%%%%%%%%%%%%%%%%%%%%%%%%%%%%%%%%%%%%%%%%%%%%%%%%%

Consider a resource allocation problem with \( m = 4 \) price levels and \( N = 20 \) resources. In this case, the cardinality of the state space is \( |\mathcal{X}| = 10626 \), highlighting the curse of dimensionality in \dyp~problems. \color{black} In our simulation, each DP iteration requires evaluating \(4 \times 10626\) state-action pairs and dealing with transition probability matrices of size \(10626 \times 4 \times 10626\). This is in addition to saving each iteration's results in a lookup table and comparing different actions \(u\) for each state \(x\) at every time step \(k\).

Moreover, from the fixed-point equation of the DP operator in the infinite-horizon setting, we have 
\[
J^* = \mathcal{R} + \alpha \, \mathcal{P}^* J^*.
\]
Solving this equation often requires matrix inversion on the order of \(10626 \times 10626\), illustrating the significant computational burden. As a result, Monte Carlo simulation is used to avoid direct matrix inversion and multiplication for such large-scale problems\color{black}. \color{black}Consider \color{black}a discount factor of \( \alpha = 0.9 \), with prices \( c = [0.8, 0.9, 1, 1.1] \), arrival probabilities \( \lambda = [0.090596, 0.048632, 0.015657, 0.005088] \), and service probabilities \( \mu = [0.483723, 0.444019, 0.024843, 0.335103] \). These probabilities \color{black} satisfy \color{black}the condition: \(
\max_{c_i, c_j \in \mathcal{U}} \sum_{x'} \big|\mathcal{P}_{xx'}(c_i) - \mathcal{P}_{xx'}(c_j)\big| < (1 - \alpha),
\) for all \( i, j = 1, \dots, m \), ensuring stability across all pairs of stationary policies (including the optimal one). \color{black} We consider $5$ features for each state $x$ of \mdp, as $\phi_1(x)= 1, \phi_i (x)=x_i,\ i= 1,\dots,\ m$, hence the parameter vector $r \in \mathbb{R}^5$. We ran $100$ Monte Carlo simulations each starting from an arbitrary initial state with the length of $50000$ iterations. In particular for Monte Carlo simulations we employ the procedure that briefly narrated in Remark \ref{monte_carlo} and Algorithm \ref{alg:off_policy_td}.

\begin{figure}[h]
	\centering
	\includegraphics[scale=0.23]{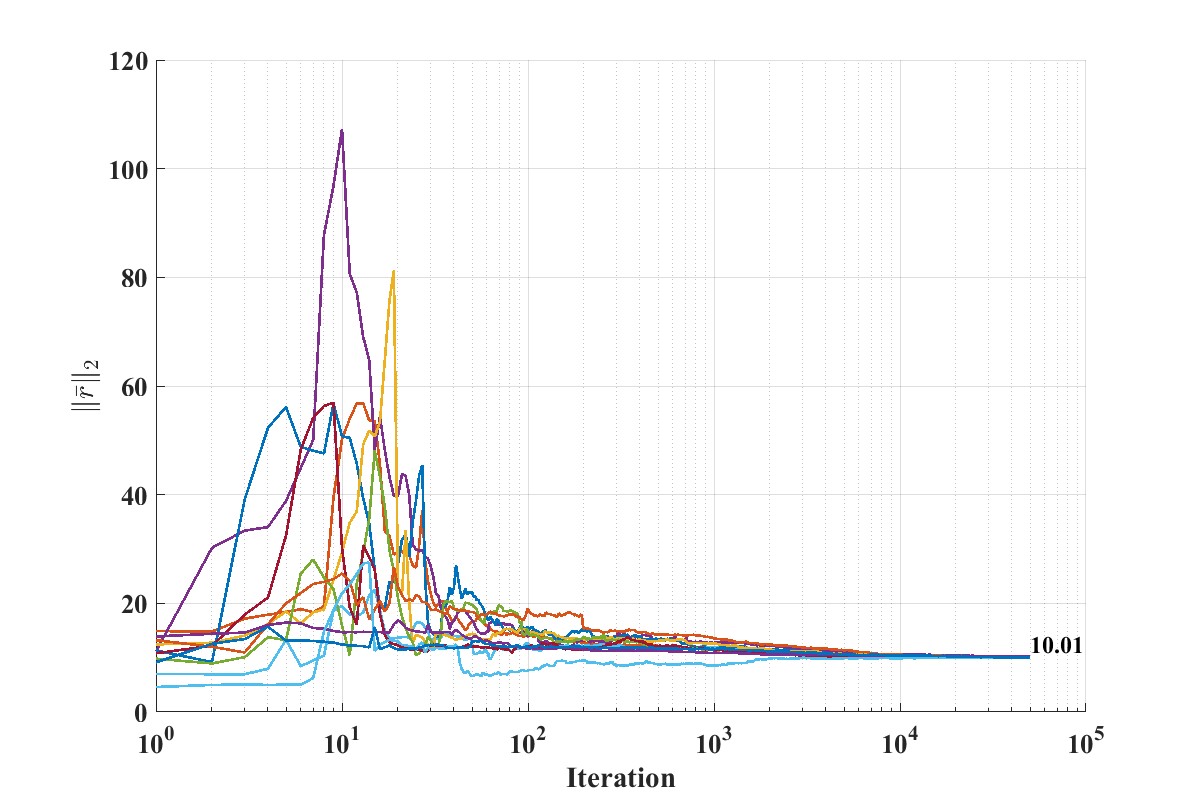}\caption{The behavior of parameter vector $r$ through off-policy \td~ approach for the resource allocation with $m=4$ prices, $N=20$ resources and $|\mathcal{X}|=10626$.}
	\label{figure_07}
\end{figure}

The results of these simulations are presented in Fig. \ref{figure_07}, where, for simplicity, we have depicted the behavior of $\big\|\bar{r}_k\big\|_2$\color{black} of parameter vectors across $10$ experiments, showing the convergence of the proposed off-policy \td~learning
%The result of these simulations are shown in Fig. \ref{figure_07} where for the sake of simplicity we only plotted the behavior of $\big\|r_k\big\|_2$ of parameter vectors in $10$ experiments which demonstrate the convergence of the proposed off-policy \td~learning 
(see Lemma \ref{c_purturbed} and Remark \ref{monte_carlo}). By averaging the parameter vectors resulted from these experiments, we have \(\bar{r}=[0.0212,\ 1.9179,\ 2.2853,\ 8.9394,\ 3.3748]^\top\).

After computing \(\bar{r}\), one can use the results of the Theorem \ref{TheoMain} to approximate $J^*$ for a given state $x$. In Fig. \ref{figure_07}, for the horizontal axis we employ logarithmic scale since the alteration rates of the curves in linear scale are not adequately detectable through iterations.

\color{black}
In the second round of the simulation, we assess the behavior of parameter vector $\bar{r}$ under different target policies as follows:
\begin{itemize}  
    \item \textcolor{black}{\textit{Greedy target policy:} The system selects the action that maximizes the instantaneous revenue, ensuring the highest immediate return at each decision step, i.e. the highest income.}
    \item \textcolor{black}{\textit{Fair target policy:} The system chooses the action that minimizes the instantaneous revenue, i.e. the lowest income.}
    \item \textcolor{black}{\textit{Random target policy:} A decision is selected randomly from the available options, introducing stochasticity into the control process to explore different state transitions.}
    \item \textcolor{black}{\textit{Hybrid Heuristic target policy:} The system uses a probabilistic approach, where with 60\% probability, it selects the decision with the lowest value of $\mu$ (potentially minimizing service rate impact), and in the remaining 40\%, it selects a random action to maintain exploration.}
    \item \textcolor{black}{\textit{Modulo-Based target policy:} The decision is determined based on the modulo operation of the sum of the current state values, enforcing a cyclic or structured selection process that balances system state transitions over time.}
\end{itemize}

\begin{figure}
    \centering
    \includegraphics[scale=0.31]{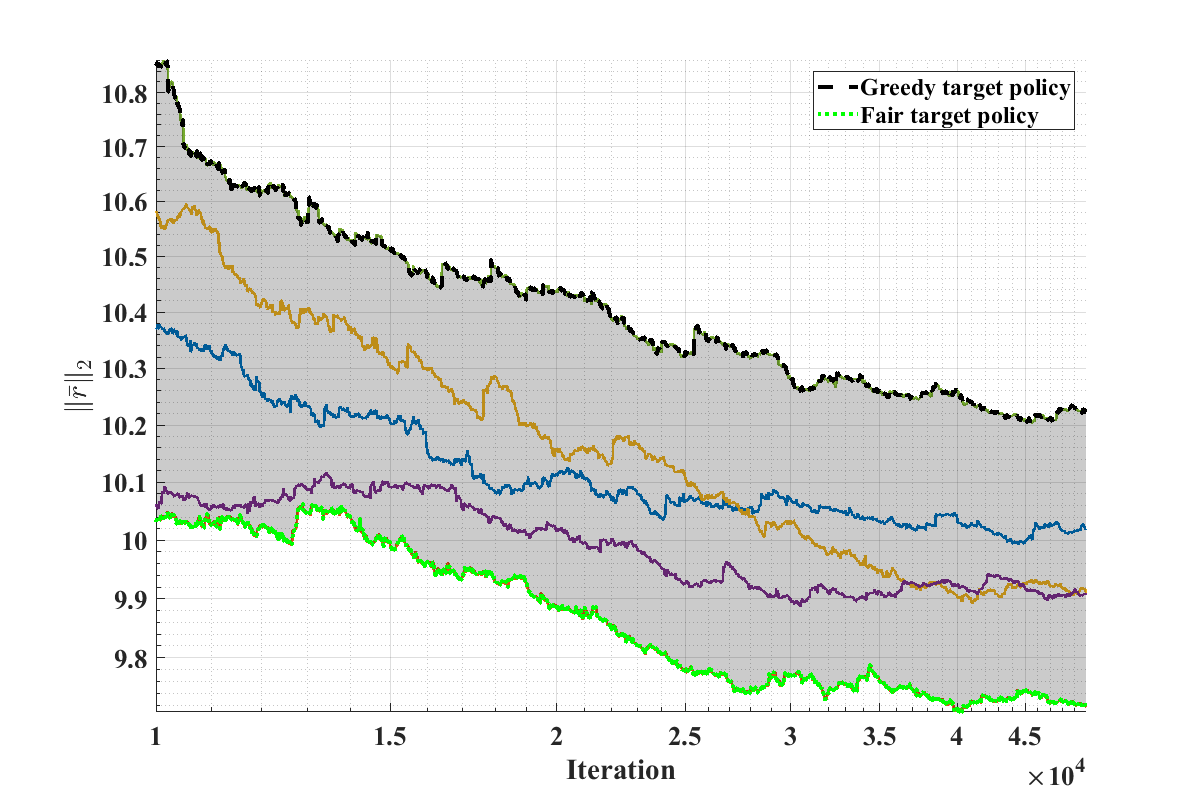}
    \caption{\textcolor{black}{Comparing different target policies on the parameter vector $\bar{r}$ through off-policy \td~approach.}}
    \label{target_policy_comparison}
\end{figure}

\color{black}{
The results of these simulations, as depicted in Fig. \ref{target_policy_comparison}, illustrate the impact of different target policies on the evolution of the parameter vector \( \bar{r} \) in the off-policy \td~learning process. The plot shows the \( \ell_2 \)-norm of \( \bar{r} \) over iterations, providing insights into the convergence behavior under varying decision-making strategies.} In particular:

\color{black}{(i) Greedy Target Policy (black dashed line): This policy consistently produces the highest parameter values throughout the iterations. Since it always selects the action that maximizes immediate revenue, it leads to a more aggressive exploitation of high-reward states, resulting in higher values of \( \|\bar{r}\|_2 \). The shaded region represents the performance bounds within which other policies operate.}

\color{black}{(ii) Fair Target Policy (green dotted line): In contrast, the Fair target policy, which minimizes immediate revenue, exhibits the lowest values of \( \|\bar{r}\|_2 \) across iterations. This outcome aligns with expectations, as this policy prioritizes decisions that yield lower short-term returns, leading to a more conservative parameter estimation.}

\color{black}{Other Policies (colored solid lines), i.e. the Random, Hybrid Heuristic, and Modulo-Based target policies, fall within the bounds set by the Greedy and Fair policies. These policies introduce varying degrees of stochasticity and structure in decision-making: the Random policy (blue) exhibits moderate behavior, as its random selection of actions balances between high and low revenue choices; the Hybrid Heuristic policy (purple) maintains a controlled exploration strategy, sometimes choosing low-service-rate decisions while preserving randomness, leading to a curve closer to the middle range; the Modulo-Based policy (brown) follows a structured cyclic decision rule, causing its trajectory to fluctuate but still remain within the expected range.}

\color{black}{Overall, the results of this simulation confirms that the Greedy policy leads to the most aggressive growth in parameter values, while the Fair policy results in the lowest values. The remaining policies provide intermediate trade-offs between exploration and exploitation.}
\color{black}

%
%The value of parameter vector

%\(\bar{r}=[0.0212, \ 1.9179, \ 2.2853, \ 8.9394, \ 3.3748]\) \color{black}

%%%%%%%%%%%%%%%%%%%%%%%%
%%%%%%%%%%%%%%%%%%%%%%%%

\section{Conclusion} \label{conclus}
In this article, we analyzed the impact of perturbations on optimal policy  in \dyp~problems and computed the resulting cost
function's deviation from its optimal value. In particular, we handled perturbations in state transition probability matrices  using an off-policy \td~projection-based approach, which also allowed to address the curse of dimensionality in large-scale~\mdp s. We also provided the necessary and sufficient conditions for the convergence of the associated Monte Carlo based simulations algorithm. Finally, we validated the presented theoretical results via a suitable numerical example.

\appendix

It is worth to highlight that a \texttt{MATLAB} package has been provided to support different implementation phase of this article which is available on
\url{https://github.com/Ali-Forootani/Off_Policy_TD_Algorithm}.
\color{black}
\bibliographystyle{ieeetr}
\bibliography{my_refs_lcss}

\begin{thebibliography}{10}

\bibitem{Forootani_IFAC}
A.~Forootani, M.~Tipaldi, M.~G. Zarch, D.~Liuzza, and L.~Glielmo, ``A
  least-squares temporal difference based method for solving resource
  allocation problems,'' {\em IFAC Journal of Systems and Control}, vol.~13,
  p.~100106, 2020.

\bibitem{Forootani_cont_sys_lett}
A.~Forootani, M.~Tipaldi, R.~Iervolino, and S.~Dey, ``Enhanced exploration
  least-squares methods for optimal stopping problems,'' {\em IEEE Control
  Systems Letters}, vol.~6, pp.~271--276, 2021.

\bibitem{Iervolino_IJC}
R.~Iervolino, M.~Tipaldi, and A.~Forootani, ``A {L}yapunov-based version of the
  value iteration algorithm formulated as a discrete-time switched affine
  system,'' {\em International Journal of Control}, vol.~96, no.~3,
  pp.~577--592, 2023.

\bibitem{lin2023policy}
Z.~Lin, J.~Ma, J.~Duan, S.~E. Li, H.~Ma, B.~Cheng, and T.~H. Lee, ``Policy
  iteration based approximate dynamic programming toward autonomous driving in
  constrained dynamic environment,'' {\em IEEE Transactions on Intelligent
  Transportation Systems}, vol.~24, no.~5, pp.~5003--5013, 2023.

\bibitem{forootani2024kernel}
A.~Forootani, R.~Iervolino, M.~Tipaldi, and S.~Baccari, ``A kernel-based
  approximate dynamic programming approach: Theory and application,'' {\em
  Automatica}, vol.~162, p.~111517, 2024.

\bibitem{Sutton_emphatic_2016}
R.~S. Sutton, A.~R. Mahmood, and M.~White, ``An emphatic approach to the
  problem of off-policy temporal-difference learning,'' {\em The Journal of
  Machine Learning Research}, vol.~17, no.~1, pp.~2603--2631, 2016.

\bibitem{sutton2008convergent}
R.~S. Sutton, C.~Szepesv{\'a}ri, and H.~R. Maei, ``A convergent {O}(n)
  algorithm for off-policy temporal-difference learning with linear function
  approximation,'' {\em Advances in neural information processing systems},
  vol.~21, no.~21, pp.~1609--1616, 2008.

\bibitem{bertsekas2011temporal}
D.~P. Bertsekas, ``Temporal difference methods for general projected
  equations,'' {\em IEEE Transactions on Automatic Control}, vol.~56, no.~9,
  pp.~2128--2139, 2011.

\bibitem{choi2006generalized}
D.~Choi and B.~Van~Roy, ``A generalized kalman filter for fixed point
  approximation and efficient temporal-difference learning,'' {\em Discrete
  Event Dynamic Systems}, vol.~16, no.~2, pp.~207--239, 2006.

\bibitem{geist2010kalman}
M.~Geist and O.~Pietquin, ``Kalman temporal differences,'' {\em Journal of
  artificial intelligence research}, vol.~39, pp.~483--532, 2010.

\bibitem{engel2005reinforcement}
Y.~Engel, S.~Mannor, and R.~Meir, ``Reinforcement learning with gaussian
  processes,'' in {\em Proceedings of the 22nd international conference on
  Machine learning}, pp.~201--208, 2005.

\bibitem{schulman2017proximal}
J.~Schulman, F.~Wolski, P.~Dhariwal, A.~Radford, and O.~Klimov, ``Proximal
  policy optimization algorithms,'' {\em arXiv preprint arXiv:1707.06347},
  2017.

\bibitem{bertsekas2023course}
D.~Bertsekas, {\em A course in Reinforcement Learning}.
\newblock Athena Scientific, 2023.

\bibitem{schulmanhigh}
J.~Schulman, P.~Moritz, S.~Levine, M.~Jordan, and P.~Abbeel, ``High-dimensional
  continuous control using generalized advantage estimation,'' {\em arXiv
  preprint arXiv:1506.02438}, 2015.

\bibitem{mnihhuman}
V.~Mnih, K.~Kavukcuoglu, D.~Silver, A.~A. Rusu, J.~Veness, M.~G. Bellemare,
  A.~Graves, M.~Riedmiller, A.~K. Fidjeland, G.~Ostrovski, {\em et~al.},
  ``Human-level control through deep reinforcement learning,'' {\em nature},
  vol.~518, no.~7540, pp.~529--533, 2015.

\bibitem{sutton2018reinforcement}
R.~S. Sutton and A.~G. Barto, {\em Reinforcement learning: An introduction}.
\newblock MIT press, 2018.

\bibitem{schulman2015trust}
J.~Schulman, S.~Levine, P.~Abbeel, M.~Jordan, and P.~Moritz, ``Trust region
  policy optimization,'' in {\em International conference on machine learning},
  pp.~1889--1897, PMLR, 2015.

\bibitem{Forootani_IET}
A.~Forootani, R.~Iervolino, and M.~Tipaldi, ``Applying unweighted least-squares
  based techniques to stochastic dynamic programming: Theory and application,''
  {\em IET Control Theory \& Applications}, vol.~13, no.~15, pp.~2387--2398,
  2019.

\bibitem{JMLRjaksch}
T.~Jaksch, R.~Ortner, and P.~Auer, ``Near-optimal regret bounds for
  reinforcement learning,'' {\em Journal of Machine Learning Research},
  vol.~11, no.~51, pp.~1563--1600, 2010.

\bibitem{gajanesliding}
P.~Gajane, R.~Ortner, and P.~Auer, ``A sliding-window algorithm for markov
  decision processes with arbitrarily changing rewards and transitions,'' {\em
  arXiv preprint arXiv:1805.10066}, 2018.

\bibitem{cheunghedging}
W.~C. Cheung, D.~Simchi-Levi, and R.~Zhu, ``Hedging the drift: Learning to
  optimize under non-stationarity,'' {\em arXiv preprint arXiv:1903.01461},
  2019.

\bibitem{cheungnon}
W.~C. Cheung, D.~Simchi-Levi, and R.~Zhu, ``Non-stationary reinforcement
  learning: The blessing of (more) optimism,'' {\em Available at SSRN 3397818},
  2019.

\bibitem{cheungreinforcement}
W.~C. Cheung, D.~Simchi-Levi, and R.~Zhu, ``Reinforcement learning for
  non-stationary markov decision processes: The blessing of (more) optimism,''
  {\em arXiv preprint arXiv:2006.14389}, 2020.

\bibitem{maonearoptimal}
W.~Mao, K.~Zhang, R.~Zhu, D.~Simchi-Levi, and T.~Başar, ``Near-optimal regret
  bounds for model-free rl in non-stationary episodic {MDP}s,'' 2020.

\bibitem{ortnervariational}
R.~Ortner, P.~Gajane, and P.~Auer, ``Variational regret bounds for
  reinforcement learning,'' in {\em Uncertainty in Artificial Intelligence},
  pp.~81--90, PMLR, 2020.

\bibitem{domingueskernel}
O.~D. Domingues, P.~M{\'e}nard, M.~Pirotta, E.~Kaufmann, and M.~Valko, ``A
  kernel-based approach to non-stationary reinforcement learning in metric
  spaces,'' {\em arXiv preprint arXiv:2007.05078}, 2020.

\bibitem{zhounonstationary}
H.~Zhou, J.~Chen, L.~R. Varshney, and A.~Jagmohan, ``Nonstationary
  reinforcement learning with linear function approximation,'' {\em arXiv
  preprint arXiv:2010.04244}, 2020.

\bibitem{touatiefficient}
A.~Touati and P.~Vincent, ``Efficient learning in non-stationary linear markov
  decision processes,'' {\em arXiv preprint arXiv:2010.12870}, 2020.

\bibitem{weinon}
C.-Y. Wei and H.~Luo, ``Non-stationary reinforcement learning without prior
  knowledge: An optimal black-box approach,'' in {\em Conference on Learning
  Theory}, pp.~4300--4354, PMLR, 2021.

\bibitem{zhongoptimistic}
H.~Zhong, Z.~Yang, Z.~Wang, and C.~Szepesv{\'a}ri, ``Optimistic policy
  optimization is provably efficient in non-stationary {MDP}s,'' {\em arXiv
  preprint arXiv:2110.08984}, 2021.

\bibitem{jinprovably}
C.~Jin, Z.~Yang, Z.~Wang, and M.~I. Jordan, ``Provably efficient reinforcement
  learning with linear function approximation,'' {\em arXiv preprint
  arXiv:1907.05388}, 2019.

\bibitem{ayoubmodel}
A.~Ayoub, Z.~Jia, C.~Szepesvari, M.~Wang, and L.~F. Yang, ``Model-based
  reinforcement learning with value-targeted regression,'' {\em arXiv preprint
  arXiv:2006.01107}, 2020.

\bibitem{zhouprovably}
D.~Zhou, J.~He, and Q.~Gu, ``Provably efficient reinforcement learning for
  discounted {MDP}s with feature mapping,'' {\em arXiv preprint
  arXiv:2006.13165}, 2020.

\bibitem{feidynamic}
Y.~Fei, Z.~Yang, Z.~Wang, and Q.~Xie, ``Dynamic regret of policy optimization
  in non-stationary environments,'' {\em arXiv preprint arXiv:2007.00148},
  2020.

\bibitem{bertsekas2019reinforcement}
D.~Bertsekas, {\em Reinforcement learning and optimal control}, vol.~1.
\newblock Athena Scientific, 2019.

\bibitem{bertsekasmultiagent}
D.~Bertsekas, ``Multiagent reinforcement learning: Rollout and policy
  iteration,'' {\em IEEE/CAA Journal of Automatica Sinica}, vol.~8, no.~2,
  pp.~249--272, 2021.

\bibitem{bertsekas2011dynamic}
D.~P. Bertsekas {\em et~al.}, ``Dynamic programming and optimal control 3rd
  edition, volume ii,'' {\em Belmont, MA: Athena Scientific}, vol.~1, 2011.

\bibitem{forootani2022transmission}
A.~Forootani, R.~Iervolino, M.~Tipaldi, and S.~Dey, ``Transmission scheduling
  for multi-process multi-sensor remote estimation via approximate dynamic
  programming,'' {\em Automatica}, vol.~136, p.~110061, 2022.

\bibitem{forootani2020least}
A.~Forootani, M.~Tipaldi, M.~G. Zarch, D.~Liuzza, and L.~Glielmo, ``A
  least-squares temporal difference based method for solving resource
  allocation problems,'' {\em IFAC Journal of Systems and Control}, vol.~13,
  p.~100106, 2020.

\bibitem{forootani2021modelling}
A.~Forootani, M.~Tipaldi, M.~Ghaniee~Zarch, D.~Liuzza, and L.~Glielmo,
  ``Modelling and solving resource allocation problems via a dynamic
  programming approach,'' {\em International Journal of Control}, vol.~94,
  no.~6, pp.~1544--1555, 2021.

\end{thebibliography}

\end{document}